\documentclass{bmcart}

\usepackage[utf8]{inputenc}
\usepackage{amsmath}
\usepackage{graphicx}
\usepackage{booktabs}
\usepackage{todonotes}
\usepackage{glossaries}
\usepackage{xspace}
\usepackage{siunitx}
\usepackage{caption}
\usepackage{subcaption}
\usepackage{bm}
\usepackage{ifthen}
\usepackage{hyperref}
\usepackage{cleveref}

\graphicspath{{./figures/}}

\makeglossaries
\newacronym{ARL}{ARL}{Adversarial Resilience Learning}
\newacronym{AI}{AI}{Artificial Intelligence}
\newacronym{DRL}{DRL}{Deep Reinforcment Learning}
\newacronym{DQN}{DQN}{Deep Q Network}
\newacronym{ENV1}{Env-1}{Environment~1}
\newacronym{ENV2}{Env-2}{Environment~2}
\newacronym{DDPG}{DDPG}{Deep Deterministic Policy Gradient}
\newacronym{A2C}{A2C}{Advantage Actor Critic}
\newacronym{TRPO}{TRPO}{Trust Region Policy Optimization}
\newacronym{KL}{KL}{Kullback-Leibler}
\newacronym{PPO}{PPO}{Proximal Policy Gradient}
\newacronym{TD3}{TD3}{Twin-Delay DDPG}
\newacronym{SAC}{SAC}{Soft Actor Critic}
\newacronym{SCADA}{SCADA}{Supervisory Control and Data Acquisition}
\newacronym{DER}{DER}{Distributed Energy Resource}
\newacronym{PV}{PV}{Photovoltaic}
\newacronym{CPS}{CPS}{Cyber-Physical System}
\newacronym{ICT}{ICT}{Information and Communication Technology}
\newacronym{CHP}{CHP}{Combined Heat and Power}
\newacronym{MV}{MV}{Medium Voltage}
\newacronym{DSO}{DSO}{Distribution System Operator}

\def\DocumentStatusSubmission{S}
\def\DocumentStatusCameraReady{C}
\def\DocumentStatus{\DocumentStatusCameraReady}
\def\dbbText{[removed for double-blind review]}
\newcommand{\dbb}[2][\dbbText]{
    \ifthenelse{\equal{\DocumentStatus}{\DocumentStatusSubmission}}{%
        #1%
    }{%
        #2%
    }%
}
\newcommand{\dbbcite}[1]{%
    \ifthenelse{\equal{\DocumentStatus}{\DocumentStatusCameraReady}}{%
        \cite{#1}%
    }{%
        \cite{dbb}%
    }%
}

\begin{document}
\begin{frontmatter}
    \begin{fmbox}
        \dochead{Research}
        \title{Learning to Attack Powergrids with DERs}
        
        \ifthenelse{\equal{\DocumentStatus}{\DocumentStatusCameraReady}}{%
            \author[
                addressref={offis},
                email={eric.veith@offis.de}
            ]{Eric MSP Veith}
            \author[
                addressref={offis},
                email={nils.wenninghoff@offis.de}
            ]{Nils Wenninghoff}
            \author[
                addressref={offis},
                email={stephan.balduin@offis.de}
            ]{Stephan Balduin}
            \author[
                addressref={uol},
                email={thomas.wolgast@uol.de}
            ]{Thomas Wolgast}
            \author[
                addressref={offis},
                email={sebastian.lehnhoff@offis.de}
            ]{Sebastian Lehnhoff}
            
            \address[id=offis]{%
                \orgname{OFFIS -- Institute for Information Technology, R\&D Division Energy},
                \street{Escherweg 2},
                \postcode{26121}
                \city{Oldenburg (Old.)},
                \cny{Germany}
            }
            \address[id=uol]{%
                \orgname{Carl von Ossietzky University Oldenburg},
                \street{Ammerländer Heerstr. 114--118},
                \postcode{26126}
                \city{Oldenburg (Old.)},
            }
        }{%
            \author[
                addressref=ue,
                email=anona1@unseen-univerity.am
            ]{Anonymous Author I}
            \author[
                addressref=ue,
                email=anona2@unseen-univerity.am
            ]{Anonymous Author II}
            \author[
                addressref=ue,
                email=anona3@unseen-univerity.am
            ]{Anonymous Author III}
            \author[
                addressref=ue,
                email=anona4@unseen-univerity.am
            ]{Anonymous Author IV}
            \author[
                addressref=ue,
                email=anona5@unseen-univerity.am
            ]{Anonymous Author V}
            \address[id=ue]{%
                \orgname{Unseen University},
                \street{Invisible Street},
                \postcode{00000}
                \city{Ankh-Morpork},
                \cny{Ankh-Morpork}
            }
        }
    \end{fmbox}
    
    \begin{abstractbox}
        \begin{abstract}
        
            In the past years, power grids have become a valuable target for cyber-attacks. Especially the attacks on the Ukrainian power grid has sparked numerous research into possible attack vectors, their extent, and possible mitigations. However, many fail to consider realistic scenarios in which time series are incorporated into simulations to reflect the transient behaviour of independent generators and consumers. Moreover, very few consider the limited sensory input of a potential attacker. In this paper, we describe a reactive power attack based on a well-understood scenario. We show that independent agents can learn to use the dynamics of the power grid against it and that the attack works even in the face of other generator and consumer nodes acting independently.
        
        \end{abstract}
    
        \begin{keyword}
            \kwd{Voltage Control}
            \kwd{Deep Reinforcement Learning}
            \kwd{Attack Vectors}
            \kwd{Distributed Energy Resources}
            \kwd{Vulnerability Analysis}
            \kwd{Cyber-Physical System}
        \end{keyword}
    \end{abstractbox}
\end{frontmatter}

\section*{Introduction}
\label{sec:introduction}

The attack on the Ukrainian power grid has shown that power grids have become valuable targets \cite{hamilton2016lights,reuters2017ukraine}. Cyber-attacks against this critical infrastructure had previously been discussed in theory, but were not considered an imminent threat yet. Nowadays, it is considered entirely possible that an attacker gains acccess to the \gls{SCADA} system---either physically or by virtually exploiting a bug that leads to a priviledge escalation---and is subsequently able to control the physical asset \cite{ei16}. This is especially important considering that power grids world-wide transition from well-controlled, central power generation and simple consumers to a distributed architecture, in which nodes become prosumers and \glspl{DER} are the norm. Not only has the number of nodes that influence the power grid increased, but also the parties involved.

A well-understood analysis of how an inverter-based generator can attack the power grid has been published by Ju and Lin~\cite{Ju2018b}. Their analysis considers \glspl{DER} as potential attack vectors, assuming that an attacker has taken over control of one or many inverters (e.\,g., wind turbines or \gls{PV} systems). Without knowledge of the grid's architecture, the authors show analytically that the attacker is able to use the reactive power control scheme of other nodes against the grid by introducing a osciallating maximum reactive power feed-in/consumption behavior. This way, an attacker can do damage of nearly twice its own reactive power feed-in/consumption capability by `leveraging' other generators.

Ju and Lin's paper is a prime example for what we call excellent first-generation analyses: They focus on an isolated issue and treat the power grid as a still object. I.\,e., while stringent in their analysis, Ju and Lin do not consider normal power grid operation: While the attacker does its work, other generators react according to their voltage control scheme, but no other nodes are active. During a normal power grid operation, other consumers and generators would still follow their schedule. In general, one cannot easily deduce from the ``petrified'' grid how the influence of time series for weather or consumption influence the effectiveness of the attack. Two scenarios are equally possible: That the attack is mitigated by other nodes as their feed-in or consumption provides enough background noise that renders the attack ineffective, or that an attack might eventually synchronize with the behavior of other nodes as to leverage at least momentarily their feed-in or consumption for the attack, too.

In this paper, we argue that a learning agent based on \gls{DRL} can learn to discover this attack. We base our research on the same premisses as Ju and Lin, namely no knowledge of the power grid structure, sensory data limited to the local attacker node only, and the same reactive power control scheme as Ju and Lin used and \cite{Ju2018b,zhu_fast_2016} introduced. However, we extend the scenario by providing a full simulation of a realistic power grid with acting independent nodes. Thus, our hypotheis is that a ``live'' power grid will foil the attack in its original, obvious fashion, but a variant of the original attack can still be discovered by a learning agent without knowledge of the rest of the grid. Hence, this paper can serve as a blue-print for a second generation of more realistic weakness analysis examples.

The remainder of this paper is structured as follows: We review the relevant literature in \nameref{sec:related-work}. The section on \nameref{sec:experiment-design} discusses the environment, sensor-actuator assignments of the attacker agent, as well as the time series and learning algorithms employed. \nameref{sec:simulation-results} shows results of our simulation and the validation of our theory. Our \nameref{sec:conclusion} also offers an outlook to future work.
\section*{Related Work}
\label{sec:related-work}

The literature that covers attacks on power systems can roughly be divided into two groups: cyber attacks and physical manipulation of the power system \dbbcite{wol21}. 

Especially cyber attacks are investigated extensively. The most important class of cyber attacks are false data injection attacks to manipulate state estimation \cite{lia17}. However, also false data injection attacks to manipulate load forcasting \cite{che19c} or automatic voltage control \cite{che19b} are possible. Further, time delay attacks, replay attacks, and theft attacks in the power system are investigated in literature \cite{meh18}. For a comprehensive review of cyber attacks refer to Mehrdad~et~al.~\cite{meh18}.

However, physical manipulation of the power system attracts more and more research interest in recent years. One or multiple compromised generators can manipulate the decentralized voltage control by leveraging the voltage control behavior of other non-attacking generators \cite{Ju2018b}. Such publications present an analytical approach that focuses on the feasibility or impact of a specific attack. However, in a previous literature survey, we found that analyzing \glspl{CPS} with learning agents is a neccessary research topic that allows to uncover new weaknesses despite the complexity of modern power grids \dbbcite{10.1145/3388218.3388222}. This is backed up by other scientist's reseach. E.\,g., Ni and Paul use Q-learning to identify the minimum number of line switchings to achieve a blackout scenario, under the assumption of knowledge about branch status \cite{ni18}. Later, they have expanded their approach to sequential attacks \cite{ni19}. \dbb{Wolgast~et.~al} demonstrate how \gls{DRL} can be used to identify unknown attack strategies that maximize profit on ancillary service markets \dbbcite{wol21}. 

While most previous publications focus on specific attack strategies or vulnerabilites---e.\,g., cyber attacks, manipulation of the powerflows, energy markets---we have previously presented an approach called \dbb{\gls{ARL}}~\dbbcite{Fischer2019} that allows the agents to explore a system while maintaining a pressure to learn through an adversarial agent. A complex co-simulation set-up including a power grid, an energy market, and a corresponding \gls{ICT} network is based on this methodology~\dbbcite{Veith2020}.

\Gls{DRL} lends itself very well for these analysis approaches, inspired by the “superhuman” performance in the game of Go. From the resurrection of (model-free) reinforcement learning with the 2013 hallmark paper by Mnih~et~al.~\cite{mnih2013playing}, to the publicly-noted achievements of AlphaGo, AlphaGo Zero, AlphaZero, and (model-based) MuZero~\cite{Lillicrap2016, Hessel2018a, Silver2016a, Silver2016, Silver2017, Schrittwieser2019}, \gls{DRL} has attracted much attention outside of the \gls{AI} domain. Much of the attention it has gained comes from the fact that, especially for Go, \gls{DRL} has found strategies better than what any human player had been able to, and this without human teaching or domain knowledge. Baker~et~al. underpin the usefulness of \gls{DRL}, specifically with competing agents, as an analysis tool: They describe how agents are able to uncover loopholes in the underlying simulation software, thus discovering a new class of strategies to achieve their goals~\cite{baker2019tooluse}. Such modern \gls{DRL}-based experimentation approaches seldomly use classic Q-learning~\cite{Hessel2018a} or A2C~\cite{mnih2016asynchronous} algorithms, but use modern variants such as \gls{SAC}~\cite{haarnoja_soft_2018} instead.

\section*{Experiment Design}
\label{sec:experiment-design}

\subsection*{Environment}

\begin{figure}
    \centering
    \includegraphics[width=0.8\textwidth]{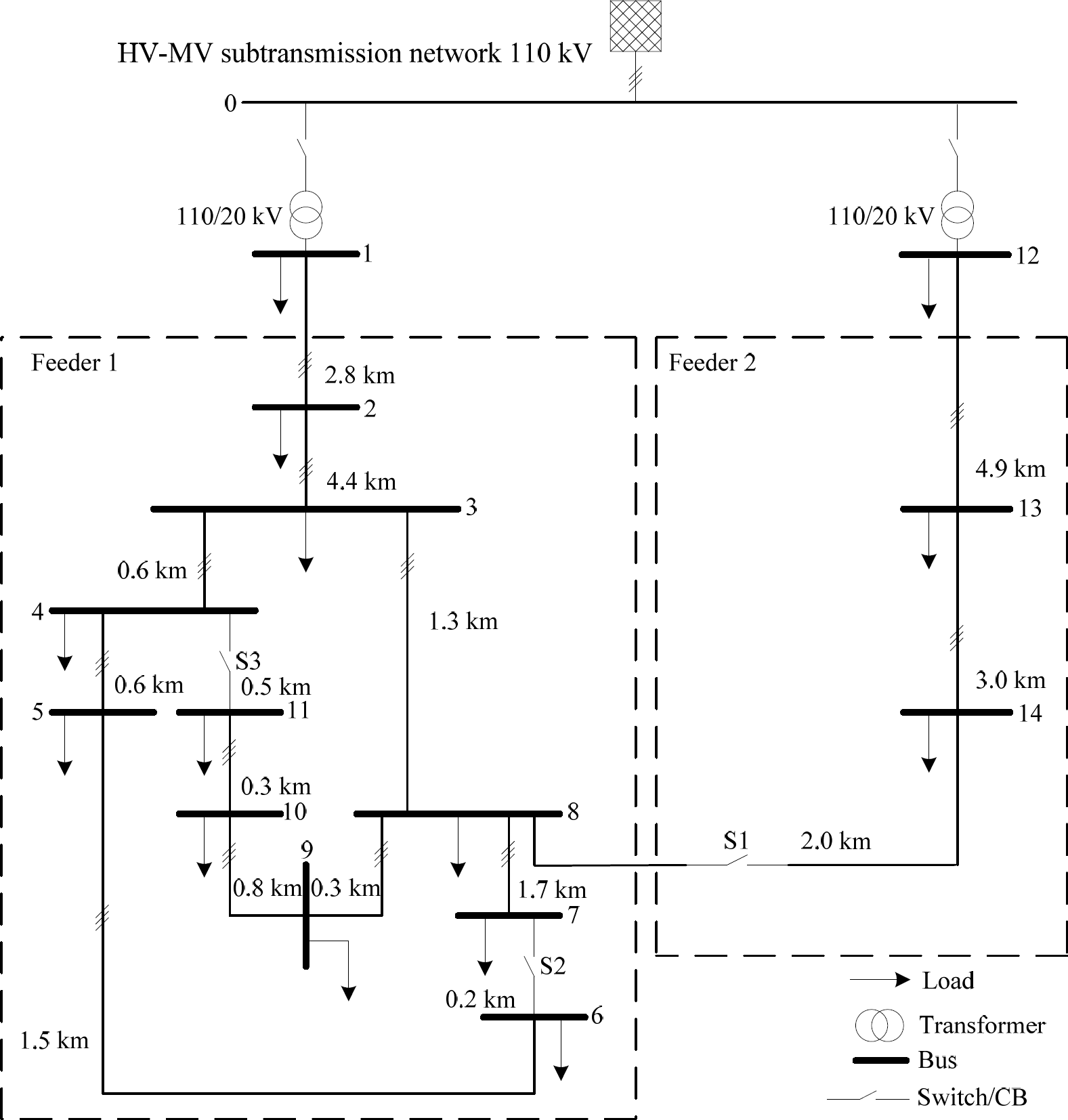}
    \caption{CIGRÉ benchmark grid used for simulation}
    \label{fig:cigre-grid}
\end{figure}

For our experiment, we employ the \dbb{MIDAS\footnote{\url{https://gitlab.com/midas-mosaik/midas}}} software suite. It incorporates the \dbb{PySimMods\footnote{\url{https://gitlab.com/midas-mosaik/pysimmods}}} software package that contains numerous models for power grid units, such as batteries, \gls{PV} or \gls{CHP} power plants. The grid model is a CIGRÉ \gls{MV} grid model \cite{1709447} as show in \cref{fig:cigre-grid}.

Each node has a constant load of \(560 + j 320\)\,kVA attached to it in order to account for this real power feed-in that occurs naturally because of the inverter model. The goal is to maintain a voltage magnitude close to \SI{1.0}{p.u.} on every bus if no action is taken. The grid has a number of \gls{PV} plants as inverter-based generators attached, as well as a super market and a small hotel that serve as consumers. These consumers follow standard load profiles according to their roles. The \gls{PV} plants' output is dependent on the solar irridation, which is governed by a multi-year weather dataset from Bremen, Germany.\footnote{Publicly available from \url{https://opendata.dwd.de/climate_environment/CDC/observations_germany/climate_urban/hourly/}.} Each PV has apparent power output of \SI{50}{MVA}. We deliberately chose this high value in order to demonstrate the attack within a short simulation time frame. For the sake of realism, one could readily assume the inverter nodes to represent wind farms. To ease modelling and without weakening the argument, we have resorted to \gls{PV} plants instead.

The attacker controls the buses~3, 4, and 8. Buses 5~to~7, 9, 11, and 13 are voltage controlled. Each benign voltage controller applies the distributed control law \cite{7361761}:

\begin{equation}\label{eq:var-control}
    \bm{q}(t+1) = \left[\bm{q}(t) - \bm{D}(\bm{V}(t)-1)\right]^+~,
\end{equation}

\noindent where the notation \([\cdot]^+\) denotes a projection of invalid values to the range \([\underline{\bm{q}^g}, \overline{\bm{q}^g}]\), i.\,e., to the feasible range of setpoints for \(q(t+1)\) of each inverter. \(\bm{D}\) is a diagonal matrix of step sizes. We have chosen a step size of 15 for each Q controller.

\subsection*{Scenarios}

The experimental validation of our hypothesis rests on the following consecutive scenarios. We define these scenarios to work from the known, working attack scenario \cite{Ju2018b} towards a scenario that implements our hypothesis. For each of the three consecutive scenarios, we change only one characteristic.

Scenarios are defined through three properties: Its \emph{goal} describes the purpose of the scenario; \emph{setup} offers a concise description of the experiment setup and simulation environment; \emph{expected result} allows to build a chain of arguments that interconnects the three scenarios.

This section contains only the description; evaluation of the actual results is offered in the \nameref{sec:simulation-results} section.

\subsubsection*{Scenario 1: Reproduction of an Original Paper}

\begin{description}

    \item [Goal] Reproduce the original paper by Ju and Lin to note the characteristics of the simulation environment and to verify the effectiveness of the originally described attack~\cite{Ju2018b}.

    \item [Setup] Consumers are not connected to the power grid. The weather provider offers weather data for a sunny summer day at noon, for each simulation step. I.\,e., the \gls{PV} plants can be operated at peak capacity.

    \item [Expected Result] The osciallating behavior of the attacking \gls{PV} plants forces the Q controllers into a similar pattern, leading to a voltage band violation.
    
\end{description}

\subsubsection*{Scenario 2: Original Attack, with Time Series}

\begin{description}

    \item [Goal] Evaluate the influence of time series data on the original attack
    
    \item [Setup] Consumers and \gls{PV} plants are connected to the grid. All nodes are influenced by the available time series data, i.\,e., the \gls{PV} plants by the weather data, the consumers by their load profiles.
    
    \item [Expected Result] The effectiveness of the original attack will depend on the current weather; the resulting data will show a clear yearly pattern.
    
\end{description}

\subsubsection*{Scenarion 3: Learn to Attack, with Time Series}

\begin{description}

    \item [Goal] Employ a learning agent to re-discover the attack described for Scenario~2.
    
    \item [Setup] As in Scenario~2. However, the attacker is now a learning agent that employs a \gls{DRL} algorithm to learn to attack the power grid.
    
    \item [Expected Result] The agent learns to create voltage band violations. We expect the learning attacker to cope better with the influence of the weather data than the original attacker.

\end{description}

\subsection*{Reward Design}

In \gls{DRL}, the design of the reward function is one of the most important steps. Our experimenting framework, \dbb{palaestrAI}, separates \emph{reward} from the \emph{objective} of an agent. This is motivated by a shortcoming of simpler \gls{DRL} experiments such as cartpole: The reward of an agent is given by the environment because of a state transition. E.\,g., for Q-learning, we commonly define the best action in a given state \(s\) as its Q value:

\begin{equation}
    Q(s,a) = r_{s,a} + \gamma \max_{a' \in \mathbf{A}} Q(s', a')~.
\end{equation}

Simple, single-agent \gls{DRL} setups assume that the reward is given by their environment and relates to the agent's goal. In multi-agent environments, this is not the case. Here, we have to distinguish between the \emph{performance of the environment} and the \emph{performance of the agent}~\dbbcite{Fischer2019}. I.\,e., if \(r_{s,a}\) is equivalent to the environment's performance when transitioning from state \(s\) given action \(a\), then the evaluation of a particular agent that considers this state transition should be denoted as \(p(r_{s,a})\).

For the simulation, we define the reward as a vectorized function of all bus voltage magnitudes after all agent actions are applied:

\begin{equation}
    \bm{r}_{s,a} = \left[ V_1, V_2, \dotsc, V_n \right]^\top~.
\end{equation}

Let \(\bm{i}_A\) denote the vector of sensors (inputs) the attacker has access to. We can then assume a function \(\mathit{filter}: (\bm{r}_{s,a}, \bm{i}_A) \mapsto \bm{r}'_{s,a}\) that reduces the vector \(\bm{r}_{s,a}\) such that it includes only those bus voltage magnitudes the attacker has sensor access to. Then, we express the objective function of the attacker as:

\begin{equation}\label{eq:objective}
    p_A(\bm{r'}_{s,a}) = 2 \left[ A \cdot \exp \left(-\frac{\left(\frac{1}{n}\sum_{r \in \bm{r'}_{s,a}}(r) - \mu\right)^2}{2\sigma^2} - C \right)\right]~.
\end{equation}

\begin{figure}
    \centering
    \includegraphics[width=\textwidth]{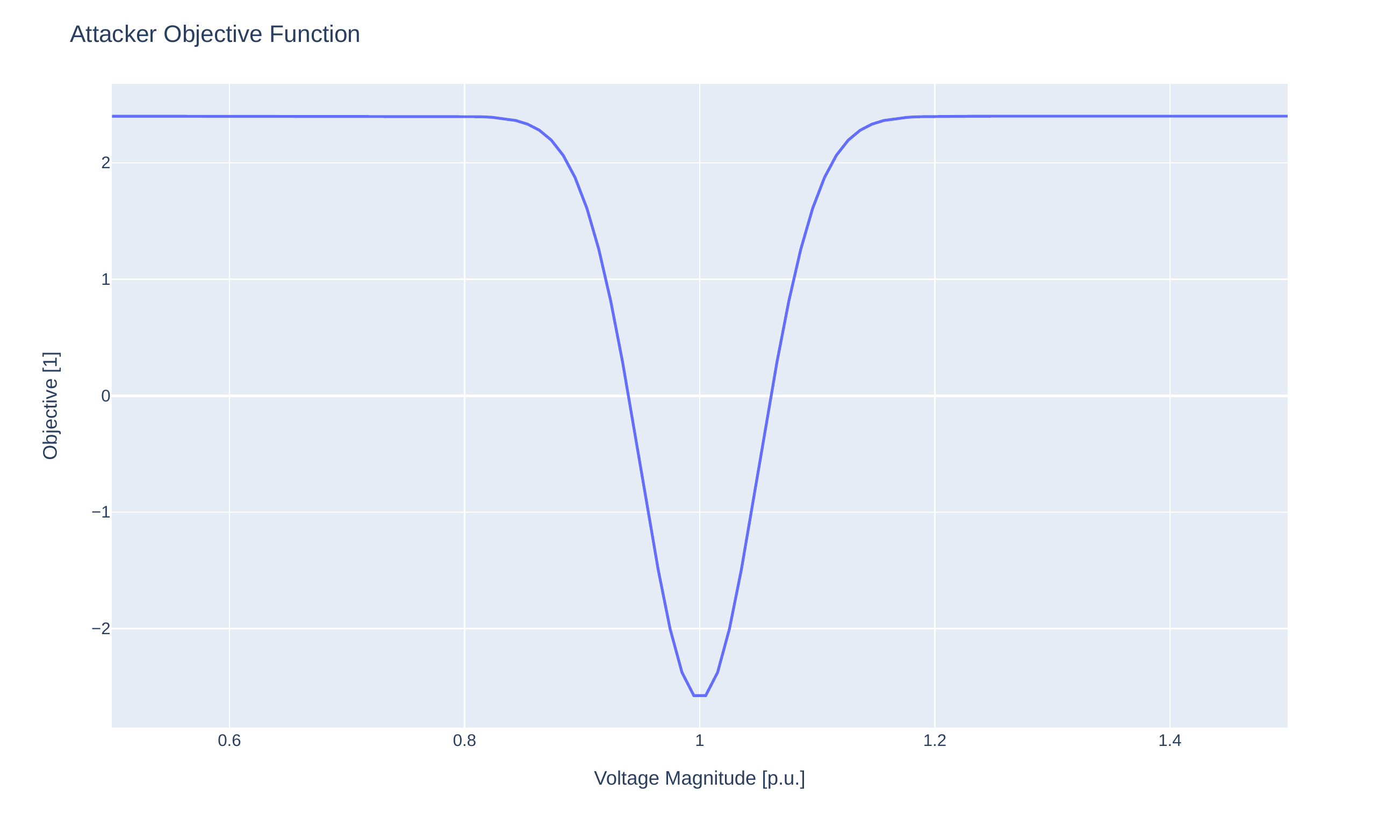}
    \caption{Objective function of the attacking agent}
    \label{fig:attacker_objective}
\end{figure}

We set \(\mu = 1\), \(\sigma = -0.05\), \(C = -1.2\), and \(A = -2.5\). The objective function is inspired by a Gaussian PDF. \Cref{fig:attacker_objective} shows that, essentially, the attacker is rewarded for a voltage band violation, while a mean voltage close to \SI{1.0}{pu} yields a negative value. This essentially encourages the attacker to force an extreme state with regards to the voltage magnitude; moreover, the influce of the weather data will drive the attacking agent towards the voltage magnitude boundares it can easily enforce.

\section*{Simulation Results}
\label{sec:simulation-results}

\subsection*{Scenario 1}

\begin{figure}[p]
    \centering
    \begin{subfigure}{\textwidth}
        \includegraphics[width=\textwidth]{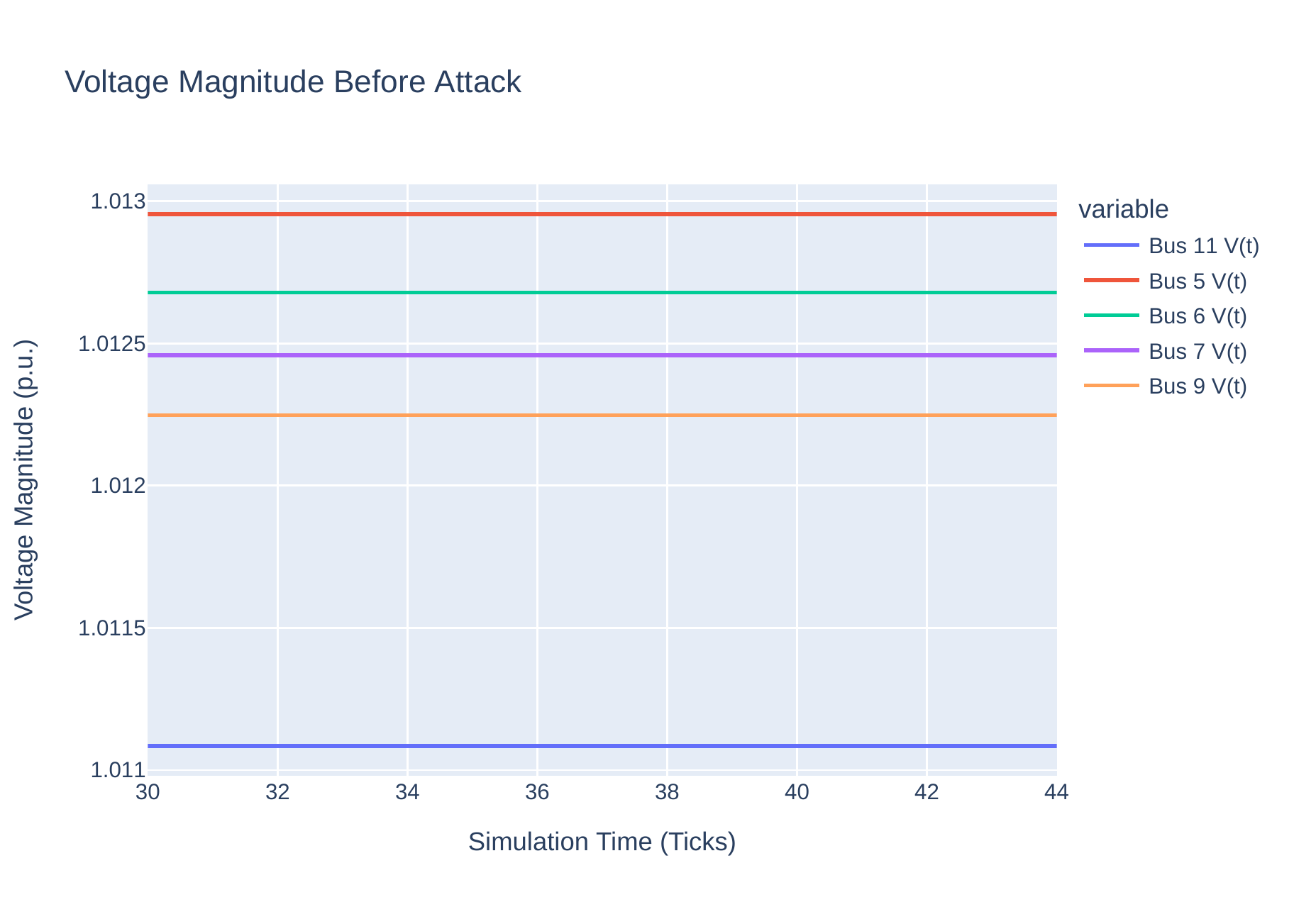}
        \caption{Voltage magnitudes on targeted buses before attack}
    \end{subfigure}
    %\hfill
    \begin{subfigure}{\textwidth}
        \includegraphics[width=\textwidth]{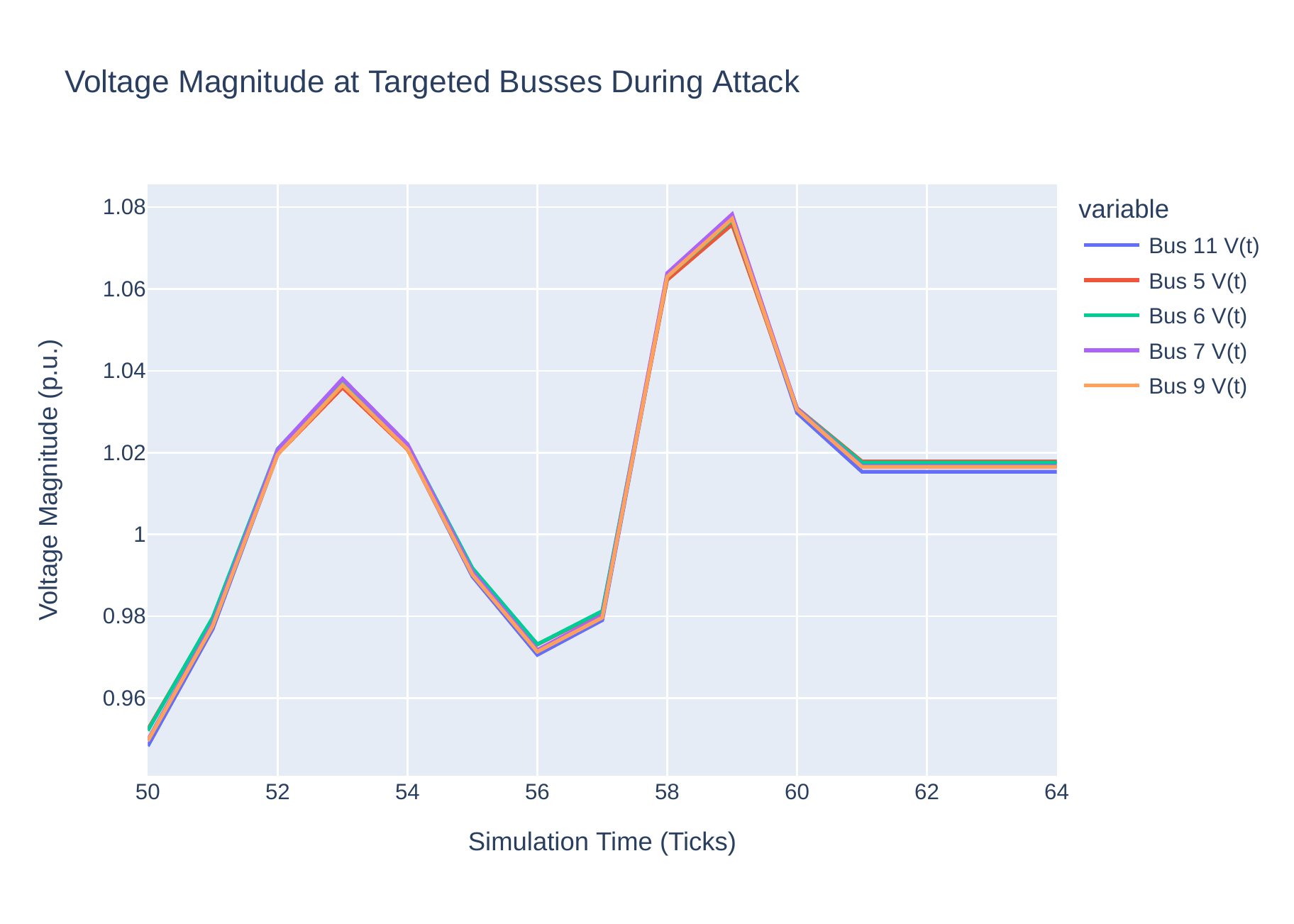}
        \caption{Voltage magnitude at benign buses during attack}
    \end{subfigure}
    \caption{System behavior for the simple alternating attacker}
    \label{fig:attack0}
\end{figure}

Scenario~1 serves as base line in order to reproduce and assertain the attack documented by Ju~and~Lin~\cite{Ju2018b}. \Cref{fig:attack0}(a) shows how the regular Volt/VAr controller within an acceptable boundary close to \SI{1.0}{pu}. \Cref{fig:attack0}(b) shows the same buses after the attacker has begun the osciallating VAr behavior. Clearly, the osciallation of the \(V(t)\) curve is visible. In addition, the second peak is higher as the first one. The attacker's behavior is always the same: It alternates between \SI{+100}{\%} and \SI{-100}{\%} of its possible reactive power injection/consumption. It confirms the general effectiveness of the attack, sans realistic time series.

\subsection*{Scenario 2}

Scenario 2 enables time series for weather and (regular) consumers, but does not change the configuration of the Volt/VAr controller or the attacker.

\begin{figure}[p]
    \centering
    \begin{subfigure}{\textwidth}
        \includegraphics[width=\textwidth]{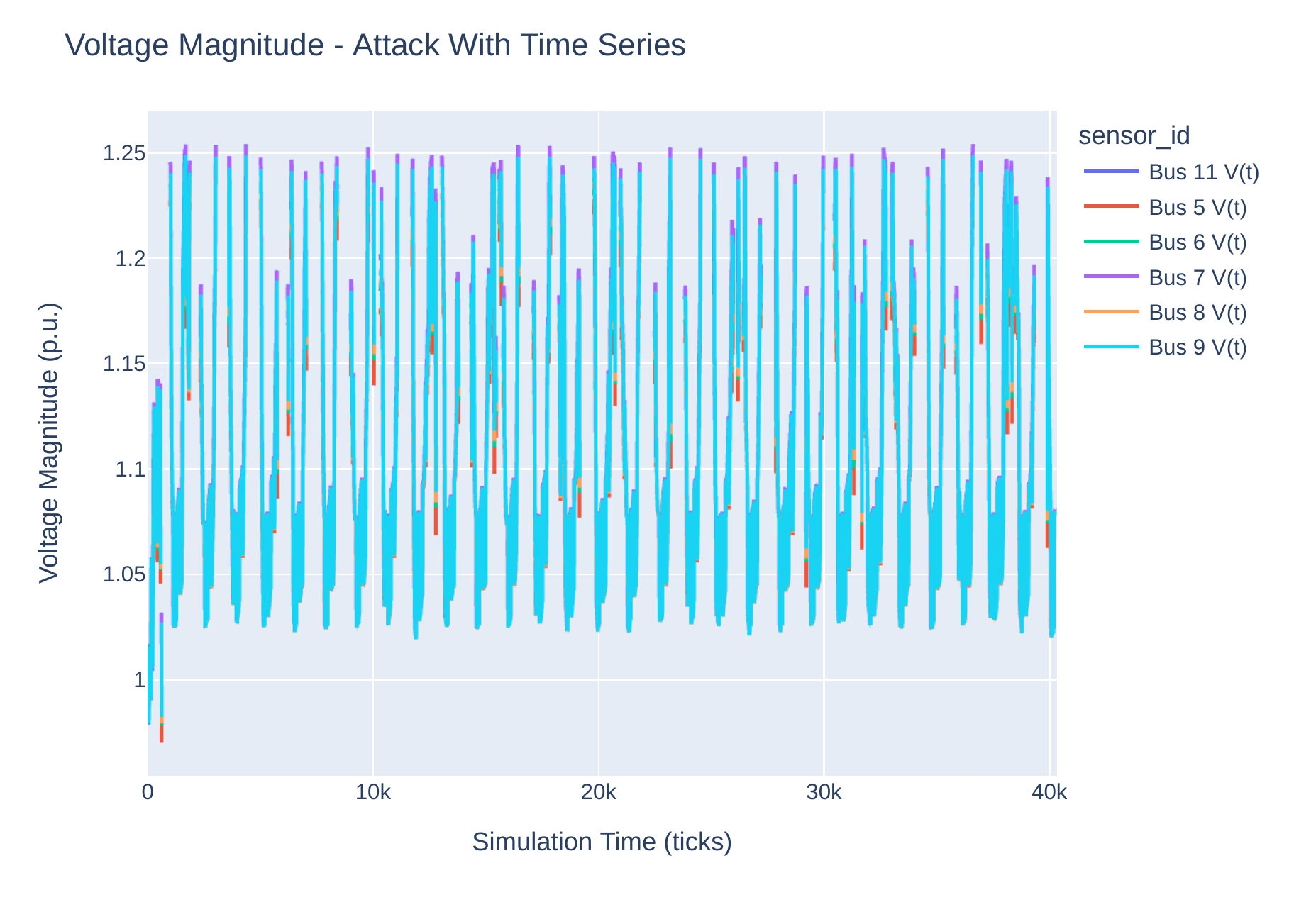}
        \caption{Voltage magnitudes}
    \end{subfigure}
    \begin{subfigure}{\textwidth}
        \includegraphics[width=\textwidth]{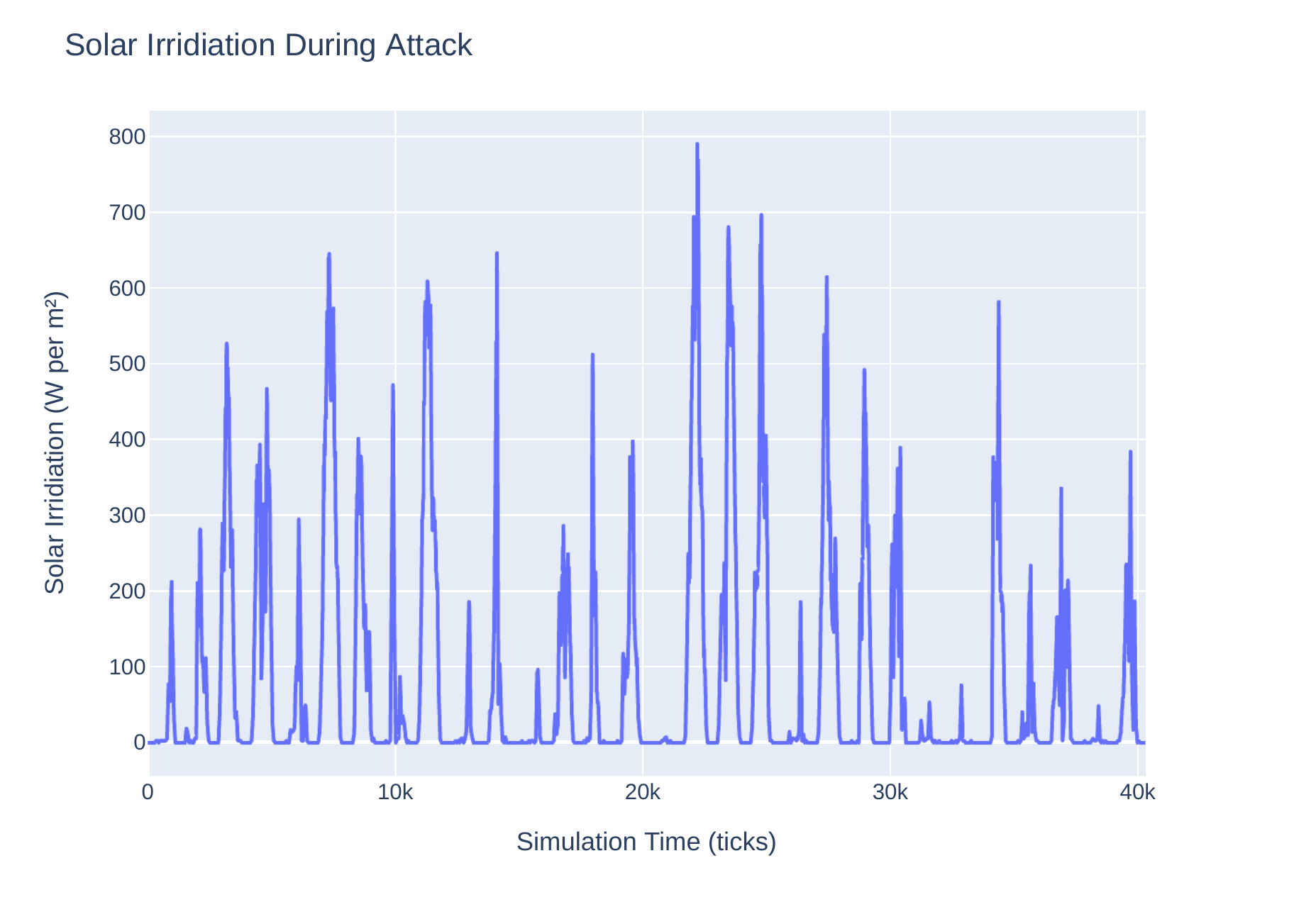}
        \caption{Solar irridiation}
    \end{subfigure}
    \caption{System behavior for the simple alternating attacker with time series}
    \label{fig:attack1}
\end{figure}

\Cref{fig:attack1} shows data from the month of July. The voltage magnitude has much higher values, even exceeding \SI{1.25}{pu}. The oscillating behavior can still be observered, but the attacker can force only 2 to 3 oscillation periods on the Volt/VAr controllers before the load flow calculation does not converge anymore.

The simulated month of July is exemplatory in showing that weather or consumer effects can lead to an increased impact of the attack. Comparing the voltage magnitude in \cref{fig:attack1}(a) with the solar irridiation plot in \cref{fig:attack1}(b), we correlate the increased voltage magnitude with the increase solar irridiation of the course of the day. Considering the distributed VAr control scheme  \cref{eq:var-control}, we see that the controller considers the current voltage magnitude as well as the previous VAr setpoint. Since the VAr setpoints are given as values relative to the inverter's current maximum, the controller overshoots the target, thereby amplifying the attack.

\subsection*{Scenario 3}

Scenario~3 introduces a learning agent as attacker in addition to scenario~3. To summarize, the third scenario then employs the CIGRÉ network with \gls{PV} feed-in, time series both for weather and for consumers, the Volt/VAr controller logic \cref{eq:var-control} defines, as well as a learning attacker. The attacker uses \gls{SAC} as learning algorithm.

\begin{figure}[p]
    \centering
    \begin{subfigure}{\textwidth}
        \includegraphics[width=\textwidth]{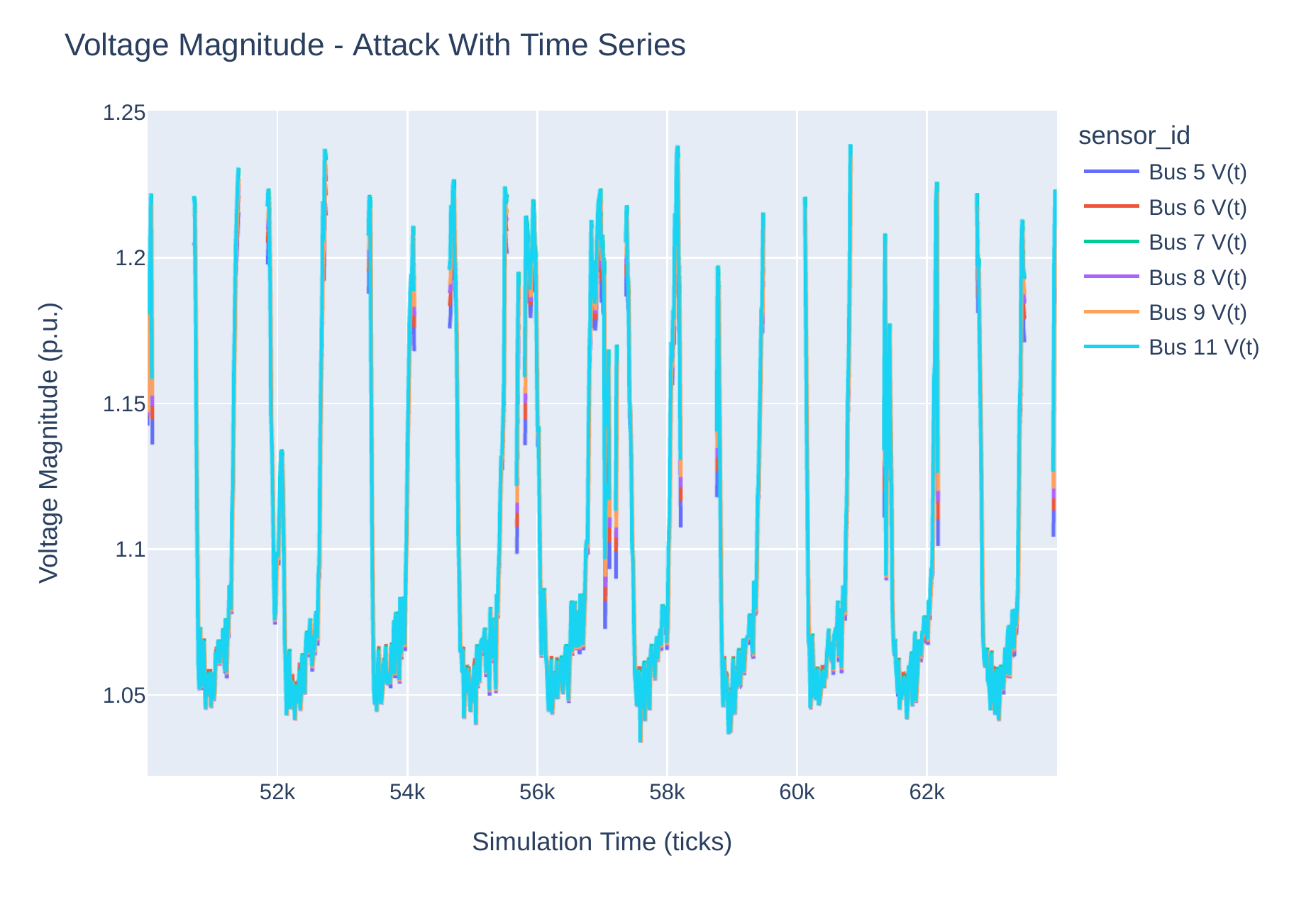}
        \caption{Voltage magnitudes}
    \end{subfigure}
    \begin{subfigure}{\textwidth}
        \includegraphics[width=\textwidth]{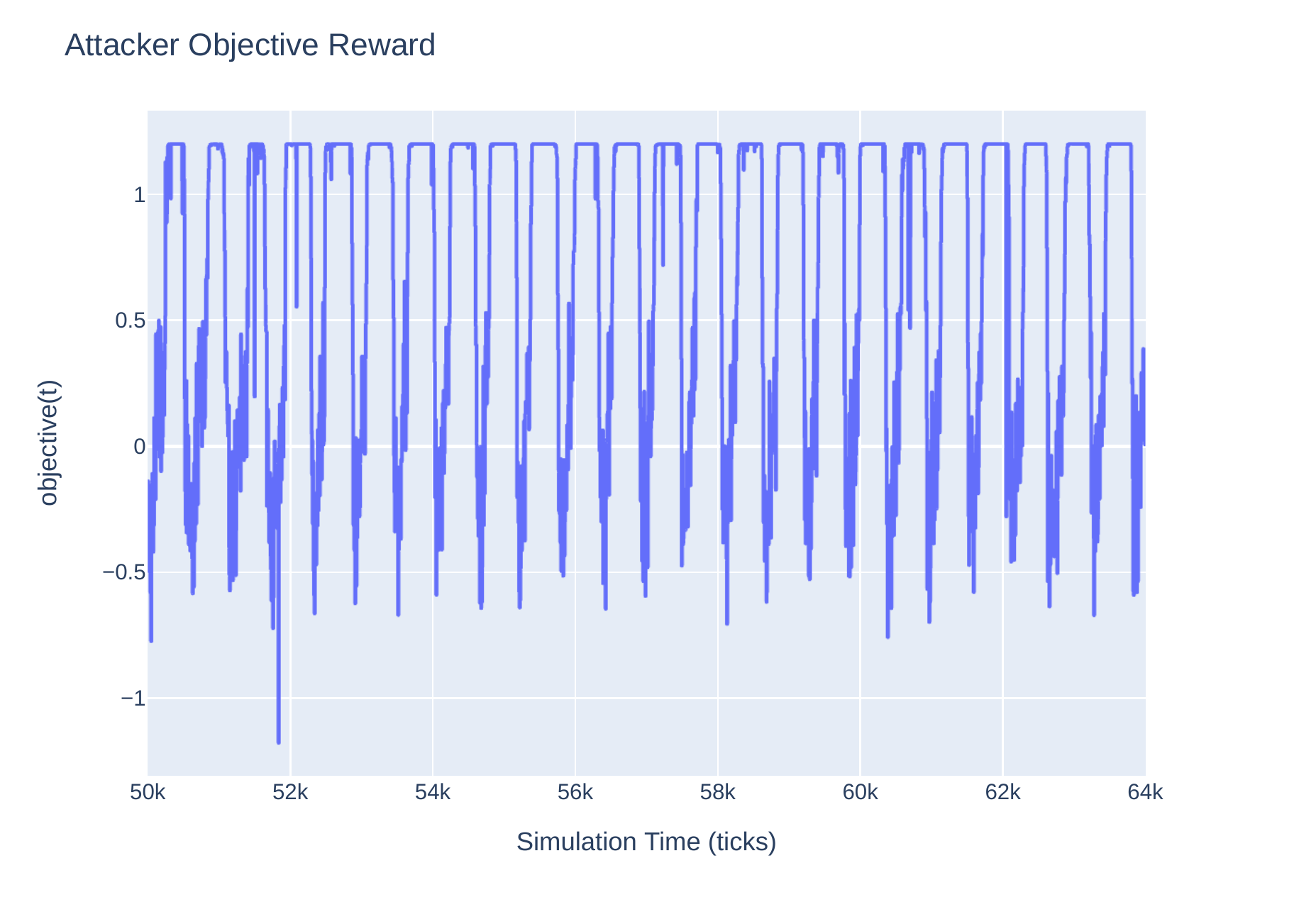}
        \caption{Solar irridiation}
    \end{subfigure}
    \caption{System behavior for the simple alternating attacker with time series}
    \label{fig:attack2}
\end{figure}

\Cref{fig:attack2}(a) confirms that the agent has learned to reproduce the attack. Note, that the agent was given no indication that the alternating setpoints behavior would lead to this result.

In fact, the denser nature of the plot in \cref{fig:attack2}(a) shows that the attacker learns to alternate the setpoints in a much quicker succession. Had the deterministic attacker in secarios~1 and~2 a holdoff time of 25~steps, none such was implemented for the \gls{SAC}-based attacker. This allowed the attacker agent to learn the value of \(\bm{D}\), i.\,e., the step size of the defender.

That the osciallating Volt/VAr control leads to the desired attack becomes evident from \cref{fig:attack2}(b). It shows the attacker's objective (cf.~\cref{fig:attacker_objective}). The objective incentivizes a voltage band violation and punishes values close to \SI{1.0}{pu}. \Cref{eq:objective} therefore does not provide an immediate incentive for the attacker to adapt the osciallating behavior; in rather disencourages it. However, in face of the Volt/VAr control regime \cref{eq:var-control}, the attacker learns that the osciallating behavior is the only sensible course of action. 

\section*{Discussion and Limitations}
\label{sec:discussion}
Volt/VAr controllers do not provide protection against malicious network actors. Deterministic Volt/VAr control schemes are vulnerable: Once known, it can be exploited in theory with any deterministic attacker. However, we show that even in absence of any knowledge about the grid topology or the control scheme, attacks are possible.

In practice, the impact of these attacks is mitigated by many other influencing factors. Usually, a power grid's design provides enough robustness against these kind of attacks. However, the vulnerability remains and can be exploited. With the general tendency to eshew physical grid extension in favor of a more intelligent grid control, the grid's inherent ``buffer'' against these kind of attacks grows ever smaller. While efficient grid control instead of physically deploying more cabling and assets is generally favorable since it is more sustainable, it also works in favor of these kinds of attacks.

The threat to the power grid depends on the degree of influence of the attacker. Our experiment design (cf. \cref{sec:experiment-design}) forms a limitation in this regard: In order to pose a consistent threat against the CIGRÉ benchmark grid, the inverter-based generators have been scaled to \SI{45}{MVA} peak output. Without loss of generability, we argue that our \gls{PV} plant model can be replaced by wind farms, where \SI{45}{MW} peak output constitute a reasonable value. 

Loads can both strengthen and weaken the deterministic attack. If loads and attackers act in the same direction at the same time, the effect is additionally amplified; if the attacker acts against the load, it weakens the attack as the attacker effectively acts like a benign Q controller. Both effects have an impact on the reaction of the voltage regulator, because it does not react to the action of the attacker, but to the bus voltage. 

Due to the design of the objective and the configuration of the power grid, the voltage only oscillates in the range above overvoltage. The objective of the agent evaluates voltages close to \SI{1.0}{pu} poorly. The grid tends to have an overvoltage due to the configuration. Undervoltage would only be possible by an action of the agent. During exploration, a sufficiently large undervoltage does not occur, so the agent only learns to generate an overvoltage. In a different network configuration, oscillation between overvoltage and undervoltage would also be possible. Adjusting the objective could also bring this about. However, this could minimize the damage, since the oscillation would have to be larger to get into critical areas. The attack could also be implemented in a network configuration with a tendency to undervoltage. 

\Gls{DRL} can reproduce the attack and exploit the Q controller's vulnerabilities. Since the Volt/VAr control scheme employed has the vulnerabilities demonstrated in scenario~1 and~2, a learning agent can find and exploit them. In addition, by exploring the solution space and adaptive properties, the agent is able to optimize the attack. 

The adaptive capability gives the attacker an unfair advantage over the defender. If the defender has a vulnerability, the attacker can search for it, find it, and exploit it, given enough time. The defender lacks an adaptive and adequate way to react. Past publications in \gls{DRL} research have shown that especially multi-agent autocurricula---our setup also being one---can find any possible solution vector and even explore and exploit weaknesses in the underlying simulation environment to this end \cite{baker2019tooluse}.

This attack shows that adaptive controllers are needed that can react depending on the situation and adapt to new attack patterns. The task of defending the power grid against attacks is more complex, since all possible attack vectors must be covered. However, in the case of learning attackers, the defender also represents a possible attack vector, as these experiments show. Complete coverage is therefore difficult to implement, due to the possible further exploration of the attacker. Instead, robustness should be pursued through continual adaptation. The continual adaptation and self-optimization of the defender reduces the risk of becoming a target of the attack itself. 

Like the work of Ju and Lin~\cite{Ju2018b}, simplifying assumptions were made in this scenario. There is no intertia in the power grid; grid codes concerning, e.\,g., voltage gradients, are also absent. A \gls{DSO} would have the attacker disconnected after the first oscillation due to violation of these codes. As future work, we will extend our scenarios with exactly these grid codes, posing the attacker to find a new attack vector.

This work shows that theoretical attack scenarios can have an impact even as the degree of realism increases. The identified vulnerabilities should be addressed in future work. However, due to the abstraction required and the still very sterile simulation environment, the threat of the identified attack vector to real power systems is low. However, it demonstrates very clearly how more realistic, complex simulations and scenarios are urgently needed to identify potential real-world vulnerabilities and develop countermeasures at an early stage. 

\section*{Conclusion}
\label{sec:conclusion}

This paper has considered voltage attacks with distributed \glspl{DER} and investigated a learning agent's ability to discover and exploit a Volt/VAr control scheme. Through scenarios that build on each other, we experimentally verified that simple oscillating behavior of a malicious agent can cause a voltage disruption by forcing other benign agents into the same oscillating behavior, thus amplifying the attack; moreover, we verified that this behavior is still relevant even with time series for consumption and weather applied to the simulation; and finally, we showed that \gls{DRL} can be used to uncover such a vulnerability.

We have seen that the \gls{DRL} agent is able to drive a far more effective attack by uncovering the step size of the Volt/VAr control scheme. We assume that an adaptive Volt/VAr controle scheme, possibly also backed by a learning agent, could mitigate the attack.

In the future, we will investigate this defender. We also plan to employ a more realistic grid by implementing grid codes and asset constraints.

\begin{backmatter}
    \printglossary
    
    \section*{Funding}
    
    \dbb{This  work  was  funded  by  the  German Federal Ministry of Education and Research through the project PYRATE (01IS19021A).}
    
    \section*{Availability of data and materials}
    
    Relevant code and experiment definitions to reproduce all experiment runs for this paper are available at the following Gitlab repository: \dbb[removed for double-blind review. All commits would uniquely identify the others. Reviewers can request the URL separately.]{\url{https://gitlab.com/Niwen/attack-paper}}

    \section*{Author's contributions}
    
    \dbb[Anonymous author I]{Eric MSP Veith} conducted experiment runs, analysis, and wrote the corresponding sections. He is also one of the main contributors to the framework \dbb{palaestrAI} that was used to carry out the analyses.
    
    \dbb[Anonymous author II]{Nils Wenninghoff} implemented the \gls{SAC} learning agent. Together with \dbb[Anonymous Author I]{Eric MSP Veith}, he created the initial reward design and was responsible for the Q control scheme. He is also one of the main contributors to the framework \dbb{palaestrAI} that was used to carry out the analyses.
    
    \dbb[Anonymous Author III]{Stephan Balduin} contributed the power grid model, models for \gls{PV}, and all time series. He is the main author of the simulation modelling framework \dbb{MIDAS}.
    
    \dbb[Anonymous Author IV]{Thomas Wolgast} sanitized the attack design and fine-tuned the scenario design.
    
    \dbb[Anonymous Author V]{Sebastian Lehnhoff} was responsbile for the overall plausibility and verification of the simulation models and testbed.

\section*{Competing interests}
  The authors declare that they have no competing interests.

    \bibliographystyle{bmc-mathphys}
    \bibliography{der-attacks}
\end{backmatter}
\end{document}